\documentclass[aps,prd,showpacs,eqsecnum,nofootinbib,twocolumn]{revtex4-1}
\usepackage{graphics}
\usepackage{graphicx}
\usepackage{pstricks}
\usepackage{epsfig}

\begin{document}
\title{Lepton number violation and neutrino masses in 3-3-1 models}
\author{Richard H. Benavides}
\affiliation{Instituto Tecnol\'ogico Metropolitano, Facultad de Ciencias Exactas y Aplicadas, Medell\'in, Colombia}

\author{Luis N. Epele, Huner Fanchiotti and Carlos Garc\'ia Canal }
\affiliation{Laboratorio de F\'\i sica Te\'orica, Departamento de 
F\'isica, Universidad Nacional de La Plata, IFLP, CONICET, C.C. 67-1900, La Plata Argentina.}



\author{William A. Ponce} 
\affiliation{Instituto de F\'\i sica, Universidad de Antioquia,
A.A. 1226, Medell\'\i n, Colombia.}
\date{\today}

\begin{abstract}
{Lepton number violation and its relation to neutrino masses is investigated in several versions of the $SU(3)_c\otimes SU(3)_L\otimes U(1)_x$ model. Spontaneous and explicit violation and conservation 
of the lepton number are considered. In one of the models (the so-called economical one), the lepton number is spontaneously violated and it is
found that the would be Majoron is not present because it is gauged away, poviding in this way the
longitudinal polarization component to a now massive gauge field.}
\end{abstract}


\maketitle

\section{\label{sec:sec1}Introduction}
The colorless and electric neutral charge distinguish the three left handed neutrinos $\nu^0_{lL}, \; l=e,\mu,\tau$ from the other fermions of the $SU(3)_c\otimes SU(2)_L\otimes U(1)_Y$ Standard Model (SM).The neutrino right handed components $\nu_{lL}^{0c}$ are not included in the spectrum of the SM which has only one scalar Higgs doublet, implying massless neutrinos at the tree level. This result holds at all orders in perturbation theory and also when non perturbative effects are taken into account due to the existence of an exact  baryon minus lepton number (B$-$L) symmetry, even if (B+L) is violated by weak sphaleron configurations. Nevertheless, neutrinos oscillate~\cite{ossc,ahm,egu,ash,nak} which implies that at least two of them have small but non zero masses.

Masses for neutrinos require physics beyond the SM connected either to the existence of $\nu_{lL}^{0c}$ and/or to the breaking 
of the (B$-$L) symmetry. If right handed neutrinos exist, the Yukawa terms leads, after electroweak symmetry breaking, 
to Dirac neutrino masses, requiring Yukawa coupling constants for neutrinos $h_\nu^\phi\leq 10^{-13}$. But $\nu_{lL}^{0c}$, singlets under the SM gauge group, can acquire large Majorana masses and turn on the see-saw mechanism~\cite{seesaw,gell,yana,moha,moha2}, an appealing and natural scenery for neutrino mass generation.

Alternatively, the left handed neutrinos, members of the SM lepton doublets  $\psi_{lL}=(\nu_l,l^-)^T_L$,  can also acquire a Majorana mass $m_\nu$ which carries weak isospin I=1 and violates lepton number L by two units, generated via non renormalizable operators of the form
$(\bar{\psi}_{lL}^c\tilde{\phi}^*)(\tilde{\phi}^\dagger\psi_{l'L})$, where $\phi=(\phi^+,\phi^0)$ is the SM Higgs doublet and $\tilde{\phi}=i\sigma_2\phi^*$. This dimension five operator is able to generate type I, type II and type III see-saw mechanisms by using for heavy fields 
an $SU(2)_L$ singlet fermion, triplet scalar and triplet fermion respectively (in this regard, see Ref. ~\cite{wein5,bon}).

The mechanism of coupling two standard model lepton doublets with a Higgs triplet $\Delta$ which in turn 
develops non zero Vacuum Expectation Values (VEV), breaking in this way the lepton number spontaneously, implies in turn  the existence of a Majoron~\cite{gelron}, particle ruled out experimentally by the $Z$ line shape measurements~\cite{lep,acco} (a singlet Majoron may still survive but with large constraints~\cite{smaj,tom}).

The variant Zee mechanism~\cite{zee,chan} can be implemented, when the L=2 Lorentz scalar $\psi_{lL}C\psi_{l^\prime L}$ (with $C$ the charge conjugation matrix) is coupled to an $SU(2)_L$ charged singlet $h^+$ with L$=-2$, introducing next 
a new scalar doublet $\phi^\prime$ and breaking the L symmetry explicitly in the scalar potential with a term of the form 
$\phi\phi^\prime h^+$. In this way, neutrino Majorana masses are generated by one loop quantum effects and the unwanted Majoron is not present.

The second Higgs doublet $\phi^\prime$ can be avoided by introducing instead a double charged Higgs singlet $k^{++}$ 
which couples to the single charged one by the trilinear coupling $k^{++}h^-h^-$ and to the right handed charged 
leptons singlets $l_R^-$ via a term of the form $l_R^-Cl^{\prime -}_Rk^{++}$, generating in this way Majorana small 
masses via two loop quantum effects by what is known as the Zee-Babu mechanism~\cite{babu,babu2}.

More examples of generation of neutrino masses via quantum effects can be found in the systematic study presented in Ref.~\cite{rzy}.

This situation motivates us to perform an extensive analysis of the lepton number symmetry
in the most relevant 3-3-1 models. In particular, we are interested in the gauging away mechanism of the Majoron that the so called economical model presents, due to the subtle connection between the lepton number generator L and one of the $SU(3)_L$ generators.

This paper is organized as follows: in Sec.~\ref{sec:sec2} we review the charge assignment and the gauge boson content of the 3-3-1 models in general, in Sec.~\ref{sec:sec3} the four possibilities for lepton number violation in the context of the minimal version of the 3-3-1 model are presented, in Sec.~\ref{sec:sec4} we classify all the 3-3-1 models without exotic electric charges and repeat the analysis presented in Sec.~\ref{sec:sec3}, but now for the so called 3-3-1 model with right-handed neutrinos. Then in Sec.~\ref{sec:sec5} we do the general analysis for the 8 different 3-3-1 models without exotic electric charges with 3 families. In Sec.~\ref{sec:sec6} the so called economical model is studied and finally, our conclusions are presented in Sec.~\ref{sec:sec7}.

\section{\label{sec:sec2}3-3-1 Models}
Some interesting extensions of the SM are based on the local gauge group $SU(3)_c\otimes SU(3)_L\otimes U(1)_x$ (3-3-1 for short) 
in which the weak sector of the SM is extended to $SU(3)_L\otimes U(1)_x$. Several models for this gauge structure have been constructed so far. 

For the 3-3-1 models, the most general electric charge operator in the extended electroweak sector is 
\begin{equation}\label{qem}
Q=a \lambda_3+ \frac{1}{\sqrt{3}}b\lambda_8 +xI_3,
\end{equation}
where $\lambda_\alpha,\;\alpha=1,2,\dots ,8$ are the Gell-Mann matrices for $SU(3)_L$ normalized as 
Tr$(\lambda_\alpha\lambda_\beta)=2\delta_{\alpha\beta}$ and $I_3=Dg(1,1,1)$ is the diagonal $3\times 3$ unit matrix. 
$a=1/2$ if one assumes that the isospin $SU(2)_L$ of the SM is entirely embedded in $SU(3)_L$; $b$ is a free parameter which defines the different possible models, and the $x$ values are obtained by anomaly cancellation. For $A_\mu^\alpha$,the 8 gauge 
fields of $SU(3)_L$, $x=0$ and thus we may write:  
\begin{equation}\label{gfi}
\sum_\alpha\lambda_\alpha A^\alpha_\mu=\sqrt{2}\left(
\begin{array}{ccc}
D^0_{1\mu} & W^+_\mu & K_\mu^{(b+1/2)} \\
W^-_\mu & D^0_{2\mu} & K_\mu^{(b-1/2)} \\
K_\mu^{-(b+1/2)} & K_\mu^{-(b-1/2)} & D^0_{3\mu} \\
\end{array}\right),
\end{equation}
where $D^0_{1\mu}=A_\mu^3/\sqrt{2}+A_\mu^8/\sqrt{6},\; 
D^0_{2\mu}=-A_\mu^3/\sqrt{2}+A_\mu^8/\sqrt{6},$ and 
$D^0_{3\mu}= -2A_\mu^8/\sqrt{6}.$  The upper indices on the gauge bosons stand for the electric charge of the particles, some of them being functions of the $b$ parameter.

\section{\label{sec:sec3}The Minimal Model}
In Ref.~\cite{pf,fram} it has been shown that, for $b=3/2$, the following fermion structure is free of all the gauge anomalies: 
$\psi_{lL}^T= (\nu_l^0,l^-,l^+)_L\sim (1,3,0),\; Q_{iL}^T=(d_i,u_i,X_i)_L\sim (3,3^*,-1/3),\; 
Q_{3L}^T=(u_3,d_3,Y)\sim (3,3,2/3)$, where $l=e,\mu,\tau$ is a family lepton index, $i=1,2$ for the first two 
quark families, and the numbers after the similarity sign mean 3-3-1 representations. The right handed fields are 
$u_{aL}^c\sim (3^*,1,-2/3),\; d_{aL}^c\sim (3^*,1,1/3),\; X_{iL}^c\sim(3^*,1,4/3)$ and $Y_L^c\sim (3^*,1,-5/3)$, 
where $a=1,2,3$ is the quark family index and there are two exotic quarks with electric charge $-4/3\; (X_i)$ and other 
with electric charge 5/3 $(Y)$. This version is called {\it minimal} in the literature, because its lepton content is just the one present in the SM.
  
For this model, the minimal scalar content required to break the symmetry, giving a realistic mass spectrum, 
consists of three triplets and one sextet: 
$\eta^T=(\eta^0,\eta_1^-,\eta_2^+)\sim (1,3,0),\; \rho^T=(\rho^+,\rho^0,\rho^{++})\sim (1,3,1),\; 
\chi^T=(\chi^-,\chi^{--},\chi^0)\sim (1,3,-1)$, and 
\begin{equation}\label{sex}
S=\left(\begin{array}{ccc}
\sigma_1^0 & s_1^+ & s_ 2^- \\
s_1^+ & s_1^{++} & \sigma_2^0 \\
s_2^- & \sigma_2^0 & s_2^{--} \\
        \end{array} \right)\sim(1,6^*,0).
\end{equation}
The scalars have Yukawa couplings to the leptons and quarks as follows:
\begin{eqnarray}
{\cal L}^l_1&=&h_{ll^\prime}^\eta\eta\psi_{lL}C\psi_{l^\prime L} + 
h_{ll^\prime}^s\psi_{lL}SC\psi_{l^\prime L} +h.c.,\\ 
{\cal L}^q_1&=&h_{ia}^uQ_{iL}^T\rho Cu^c_{aL}
+h_{ia}^dQ_{iL}^T\eta Cd^c_{aL}\\ \nonumber 
&+&h_{ij}^XQ_{iL}^T\chi CX^c_{jL} 
+h_{3a}^dQ_{3L}^T\rho^* Cd^c_{aL}\\ \nonumber 
&+& h_{3a}^uQ_{3L}^T\eta^* Cu^c_{aL}+h^YQ_{3L}^T\chi^* CY^c_{L} + h.c.,
\end{eqnarray}
with vacuum expectation values (VEV) given by $\langle\eta^0\rangle=v_1, \; \langle\rho^0\rangle=v_2,\; 
\langle\chi^0\rangle=v_3,\; \langle\sigma_1^0\rangle=v_4$ and $\langle\sigma_2^0\rangle=v_4^\prime$.

One of the main characteristics of this model is the fact that the lepton number L is not a good quantum number because both, 
the charged lepton and its antiparticle are in the same multiplet; as a consequence, L does not commute with the 
electroweak extended gauge symmetry.

The assignment of L starts with the SM assignments~\cite{ng} 

\begin{eqnarray*}{\rm L}&& (l^-_L,\; \nu_{lL})=-{\rm L}(l^+_L)=1,\\ 
 {\rm L}&&(u_{aL},\; u_{aL}^c,\; d_{aL},\; d_{aL}^c,\; W^\pm_\mu ,\; D_{1\mu}^0 ,\; D_{2\mu}^0,\; D_{3\mu}^0)=0;
\end{eqnarray*} 
then, looking to the Yukawa interactions of the SM particles and imposing L=0 in the covariant derivative implies 
\begin{eqnarray*}{\rm L}&&(K^{++},\; K^+,\; Y_L,\; X_{iL}^c)=-2\\
{\rm L}&&(K^{--},\; K^-,\; X_{iL},\; Y_L^c)=2.
\end{eqnarray*} 
For the scalars, L is assigned by inspection of the Yukawa coupling constants and one finds   
\begin{eqnarray*}{\rm L}&&(\chi^-,\;\chi^{--},
\; s_2^{--})=2, \\
 {\rm L}&&(\eta_2^+,\; \rho^{++},\; \sigma_1^0,\; s_1^+,\; s_1^{++})=-2,\\
 {\rm L}&&(\eta^0,\;\eta_1^-,\; 
\chi^0,\; \rho^+,\; \rho^0,\;  \sigma_2^0,\; s_2^-)=0.\end{eqnarray*} 
Notice that $X_i$ and $Y$ are bi-leptoquarks and 
$K^+,\; K^-,\; K^{++}$ and $K^{--}$ are bi-lepton gauge bosons. Finally, the physical gauge bosons related to the neutral 
currents of the model have L=0.

It is interesting to notice that the above lepton numbers of the individual components of each multiplet can be written as~\cite{tully}
\begin{equation}\label{lll}
{\rm L}=\frac{2\lambda_8}{\sqrt{3}}+{\cal L}I_3,
\end{equation}
where ${\cal L}$ is a global symmetry of the Lagrangian which is not broken by the VEV, and is related to the following assignment: ${\cal L}(\psi_{lL})=1/3,\; 
{\cal L}(Q_{iL})=2/3,\; {\cal L}(Q_{3L},\;\eta ,\; S,\; \rho)=-2/3,\;
{\cal L}(\chi)=4/3,\;{\cal L}(X^c_{iL})=-2,\; {\cal L}(Y_L^c)=2,$ and ${\cal L}(u_a^c,\;d_a^c,\; A_{\alpha\mu})=0$.

The former analysis shows that since L$(\eta^0,\; \chi^0,\; \rho^0,\; \sigma_2^0)=0$, the only place where the L number 
can be spontaneously violated is in $\sigma_1^0$, but it may be explicitly violated in the scalar potential. As a matter 
of fact, a term like 
\begin{eqnarray}\nonumber
V_{LV}&=&f_1\eta S\eta+f_2SSS+\kappa_1(\chi^\dagger\eta)(\rho^\dagger \eta)+ 
\kappa_2\eta^\dagger S\chi\rho \\ \label{vlv}
&+&\kappa_3\chi\rho SS + h.c.,
\end{eqnarray}
explicitly violates $\Delta {\cal L}=\Delta$L=$\pm 2$ when all the VEV are zero, leaving $\lambda_8$ unbroken. Then, the four possibilities of lepton number violation in the context of this model are thus:

(1) $V_{LV}=0$ and $\langle S\rangle=0$. This is the minimal 3-3-1 Pisano-Pleitez-Frampton model where total lepton number is conserved and neutrinos are massless particles. Consecuently, this version of the model is in conflict with the existence of massive neutrinos.

(2) $V_{LV}=0$ but $\langle\sigma_1^0\rangle\neq 0$. In this case the lepton number is spontaneously broken leading to a triplet Majoron. This case has been analyzed in Ref~\cite{pgm}.

(3) $V_{LV}\neq 0$ and $\langle\sigma_1^0\rangle=0$. L is violated explicitly and non zero masses for neutrinos can be 
generated from quantum corrections.

(4) The case for $V_{LV}\neq 0$ and $\langle\sigma_1^0\rangle\neq 0$ is also possible, with a rich phenomenology which may 
include a light pseudo Goldstone Majoron~\cite{llw}.

\section{\label{sec:sec4}3-3-1 Models Without Exotic Electric Charges}
If one wishes to avoid exotic electric charges as the ones present in the minimal model, one must choose $b=1/2$, in Eq. 
(\ref{qem}). Following \cite{juan,ander} we can find six sets of fermions which contain the antiparticles of the charged particles 
which are  
\begin{itemize}
\item $S_1=[(\nu^0_\alpha,\alpha^-,E_\alpha^-);\alpha^+;E_\alpha^+]_L$ with quantum numbers $(1,3,-2/3);(1,1,1)$ and $(1,1,1)$ respectively.
\item $S_2=[(\alpha^-,\nu_\alpha,N_\alpha^0);\alpha^+]_L$ with quantum numbers $(1,3^*,-1/3)$ and $(1,1,1)$ respectively.
\item $S_3=[(d,u,U);u^c;d^c;U^c]_L$ with quantum numbers $(3,3^*,1/3);\; (3^*,1,-2/3);\; (3^*,1,1/3)$ and $(3^*,1,-2/3)$ respectively.
\item $S_4=[(u,d,D);u^c;d^c;D^c]_L$ with quantum numbers $(3,3,0);\; (3^*,1,-2/3);\; (3^*,1,1/3)$ and $(3^*,1,1/3)$ respectively.
\item $S_5=[(e^-,\nu_e,N_1^0);(E^-,N_2^0,N_3^0);(N_4^0, E^+,e^+)]_L$ with quantum numbers $(1,3^*,-1/3)$;$(1,3^*,-1/3)$ and $(1,3^*,2/3)$ respectively.
\item $S_6=[(\nu_e, e^-,E_1^-);(E^+_2,N_1^0,N_2^0);(N_3^0, E^-_2,E_3^-)$;  $e^+; E_1^+; E_3^+]_L$ with quantum numbers $(1,3,-2/3)$; $(1,3,1/3)$; $(1,3,-2/3)$; $(111), (111)$;  and $(111)$ respectively.
\end{itemize}

The different anomalies for these six sets are \cite{juan} found in Table I.

\begin{table}[here!]{{TABLE I.\\Anomalies for 3-3-1 fermion fields structures}}\label{tabl1}

\begin{tabular}{||l|cccccc||}\hline\hline
Anomalies & $S_1$ & $S_2$ & $S_3$ & $S_4$ & $S_5$ & $S_6$ \\ \hline
$[SU(3)_C]^2U(1)_X$ & 0 & 0 & 0 & 0 & 0 & 0 \\
$[SU(3)_L]^2U(1)_X$ & $-2/3$  & $-1/3$ & 1 & 0& 0 & -1\\
$[Grav]^2U(1)_X$ & 0 & 0 & 0 & 0 & 0 & 0 \\
$[U(1)_X]^3$ & 10/9 & 8/9 & $-12/9$ & $-6/9$& 6/9& 12/9 \\
$[SU(3)_L]^3$ & 1 & $-1$ & $-3$ & 3 & $-3$ & 3\\
\hline\hline
\end{tabular} 
\end{table}
 
With this table, anomaly-free models, without exotic electric
charges can be constructed for one, two or more families. 

As noted in Ref. \cite{juan}, there are eight three-family models that are anomaly free, which are:

\begin{itemize}
 \item Model A: with right-handed neutrinos\\
$3S_2+S_3+2S_4$.
\item Model B: with exotic electrons\\
$3S_1+2S_3+S_4$
\item Model C: with unique lepton generation one (three different lepton families)\\
$S_1+S_2+S_3+2S_4+S_5$
\item Model D: with unique lepton generation two\\
$S_1+S_2+2S_3+S_4+S_6$
\item Model E: hybrid one (two different lepton structures)\\
$S_3+2S_4+2S_5+S_6$
\item Model F: hybrid two\\
$2S_3+S_4+S_5+2S_6$
\item Model G: carbon copy one (three identical families as in the SM)\\
$3(S_4+S_5)$
\item Model H: carbon copy two\\
$3(S_3+S_6)$
\end{itemize}

\subsection{\label{sec:sec41}The 3-3-1 model with right-handed neutrinos}
Introduced in Ref.~\cite{long,foot}, it has the following 3-3-1 anomaly free fermion structure:
\begin{eqnarray*}
\psi_{lL}^T & = & (l^-,\nu_l^0,N_l^{0})_L \sim (1,3^*,-1/3),\; l_L^+\sim(1,1,1), \\   
Q_{iL}^T & = & (u_i,d_i,D_i)_L  \sim  (3,3,0), \\
 Q_{3L}^T & = & (d_3,u_3,U)_L  \sim  (3,3^*,1/3),
\end{eqnarray*}
where $l=e,\mu,\tau$ is a family lepton index, $N_{lL}^{0}$ stands for 
electrically neutral Weyl state, and $i=1,2$ for the first two quark families. The right handed quark fields are
\begin{eqnarray*}
u_{aL}^c &\sim& (3^*,1,-2/3),\; \; d_{aL}^c\sim (3^*,1,1/3),\\ 
D_{iL}^c&\sim&(3^*,1,1/3),\; \; U_L^c\sim (3^*,1,-2/3),
\end{eqnarray*} 
where again $a=1,2,3$ is the quark family index and there are two exotic quarks with electric charge $-1/3\; (D_i)$ 
and other with electric charge 2/3 $(U)$. 

The minimal scalar content required to break the symmetry, giving a realistic mass spectrum, consists now of only three 
triplets~\cite{long,foot}: 
\begin{eqnarray}\label{higg}\nonumber
\rho^T&=&(\rho^0_1,\rho^+_2,\rho^{+}_3)\sim (1,3^*,2/3),\\  \nonumber 
 \eta^T&=&(\eta_1^-,\eta^0_2,\eta_3^0)\sim (1,3^*,-1/3),\\
 \chi^T&=&(\chi_1^-,\chi_2^0,\chi_3^0)\sim(1,3^*,-1/3),
\end{eqnarray} 
with VEV given by $\langle\rho^0\rangle^T=(v_1,0,0),\; \langle\eta^0\rangle^T=(0,v_2,0),$ and $\langle\chi^0\rangle^T=(0,0,V)$.

A careful analysis of the Yukawa terms for the lepton sector 
\begin{eqnarray}\label{llep}
 {\cal L}^Y_{lep}=h^e_{ll^\prime}\rho^* \psi_{lL}Cl_L^{+\prime} + h_{ll^\prime}\rho \psi_{lL}C\psi_{l^\prime L},
\end{eqnarray}
shows that $N_{lL}^0$, the third component of the fermion triplet $(1,3^*,-1/3)$, must be identified with $\nu^{0c}_{lL}$,
the antiparticle of $\nu^0_{lL}$. As a consequence, L is not a good quantum number in the context of this model, because
L does not commute with the symmetry $SU(3)_L\otimes U(1)_X$.

Doing a similar analysis that the one presented for the minimal model, we obtain the following lepton number 
assignments~\cite{chang} 
\begin{eqnarray} \label{lnass}
{\rm L} &&(l^-_L,\; \nu^0_{lL})=-{\rm L} (l^+_L,\;\nu^{0c}_{lL})=1, \\ \nonumber
 {\rm L} &&(u_{aL},\; u_{aL}^c,\; d_{aL},\; d_{aL}^c,\; W^\pm_\mu ,
 \; D_{1\mu}^0 ,\; D_{2\mu}^0,\; D_{3\mu}^0 )=0;\\ \nonumber 
 {\rm L} &&(K^{+},\; K^0,\; U_L,\; D_{iL}^c)=-{\rm L} (K^-,\; \overline{K}^0,\; D_{iL},\; U_L^c)=-2,\\ \nonumber
{\rm L} &&(\chi_1^-,\;\chi^0_2)=-{\rm L} (\rho_3^+,\; \eta_3^0)=2, \\ \nonumber
{\rm L}&&(\rho_1^0,\; \rho_2^+,\; \eta_1^-,\;\eta_ 2^0,\;\chi_3^0)=0.
\end{eqnarray}

Notice that the new quarks $D_i$ and $U$ are bileptoquarks and $K^+,\; K^-,\; K^0$ and $\overline{K}^0$ are bi lepton 
gauge bosons. 

Again, Eq. (\ref{lll}) can be used to write the previous lepton number assignment using now the following ${\cal L}$ values:
${\cal L}(\psi_{lL})=1/3,\; 
{\cal L}(Q_{iL})=2/3,\; {\cal L}(Q_{3L},\;\eta ,\;\rho)=-2/3,\;
{\cal L}(\chi)=4/3,\;{\cal L}(D^c_{iL})=-2,\; {\cal L}(U_L^c)=2, \; {\cal L}(l^+_L)=-1$,  
and ${\cal L}(u_{aL}^c,\;d_{aL}^c,\; A_{\alpha\mu})=0$, values in agreement with the one presented in Ref.~\cite{chang}.

For this model, the quark mass spectrum was analyzed in Ref.~\cite{chang}, using only the following lepton number conservation Yukawa potential:
\begin{eqnarray}\label{lylnc} 
{\cal L}^Y_{LNC}&=&h^U \chi^*Q_{3L}CU_L^c+h^D_{ij}\chi Q_{iL}CD^c_{jL}\\ \nonumber
&&+h^d_a\rho^*Q_{3L}Cd^c_{aL}+h^u_{ia}\rho Q_{iL}Cu_{aL}^c\\ \nonumber
 && +h_{3a} \eta^* Q_{3L}Cu_{aL}^c+h_{ia}\eta Q_{iL}Cd^c_{aL}+ h.c.,
\end{eqnarray}
which conserves both the global numbers  L and $\mathcal{L}$. But the most general Yukawa potential for quarks must also 
include the following terms 
 
\begin{eqnarray}\label{lylnv}
{\cal L}^Y_{LNV}&=& h_a^u\chi^*Q_{3L}Cu_{aL}^c + h^d_{ia}\chi Q_{iL}Cd_{aL}^c \\ \nonumber
&&+h_i^D\rho^*Q_{3L}CD_{iL}^c + h^U_{i}\rho Q_{iL}CU_{L}^c \\ \nonumber
&&+ h^U\eta^*Q_{3L}CU_{L}^c + h^D_{ij}\eta Q_{iL}CD_{jL}^c + h.c., \\ \nonumber
\end{eqnarray}
which explicitly violates the global numbers L and $\mathcal{L}$. This avoids the possible existence of a Majoron in the 
context of this model.

The Yukawa Lagrangian for the neutral leptons extracted from (\ref{llep}), and in the basis 
$(\nu_1,\nu_2,\nu_3,\nu_{1}^c,\nu_{2}^c,\nu_{3}^c)$, 
produces the following tree level neutrino mass matrix

\begin{equation}
M=\left(\begin{array}{cccccc}
0 & 0 & 0 & 0  & a  & b   \\
0 & 0 & 0 &-a & 0 & c \\
0 & 0  & 0  & -b & -c & 0  \\
0 & -a & -b & 0 & 0 & 0  \\
a & 0 & -c & 0 & 0 & 0  \\
b & c & 0 & 0 & 0 & 0  
        \end{array} \right),
\end{equation}
where the entries are Dirac masses at the SM scale, times Yukawa couplings, with eigenvalues $(0,0,\pm m_\nu,\pm m_\nu)$ where 
$m_\nu = \sqrt{a^2+b^2+c^2}$ which stands for three Dirac neutrinos, one massless and two degenerated. The model is viable 
only for very small Yukawa couplings constants and radiative corrections able to remove the degeneracies (analysis done to a limited extent in Ref.~\cite{chang}).      

In general $\chi_2^0$ and $\eta_3^0$ can have a VEV different from zero which could imply spontaneous symmetry breaking 
of the lepton number L. But L can also be broken explicitly in the scalar potential by terms of the form 
\begin{eqnarray}\nonumber
V_{LV}^\prime &=& \mu\chi^\dagger\eta + \eta^\dagger\chi(\kappa_1|\rho|^2+\kappa_2|\eta|^2+\kappa_3|\chi|^2)\\
&+&\kappa_4|\chi^\dagger\eta|^2+\kappa_5(\eta^\dagger\rho)(\rho^\dagger\chi) + h.c.,
\end{eqnarray}
which again satisfy $\Delta{\cal L}=\Delta L=\pm 2$ when all the VEV are zero, leaving $\lambda_8$ to be broken explicitly.

As in the minimal model there are four different cases:\\
(1) $V_{VL}^\prime=0,\; \langle\chi_2^0\rangle=\langle\eta_3^0\rangle=0$. The total lepton number is conserved and the neutrinos can pick up only Dirac type masses.\\
(2) $V_{VL}^\prime=0,\; \langle\chi_2^0\rangle\neq 0$ and/or $\langle\eta_3^0\rangle \neq 0$. The lepton number L is now spontaneously violated. This case has been analyzed in Ref.~\cite{pires} where a CP odd Majoron was found.\\
(3) $V_{VL}^\prime\neq 0,\; \langle\chi_2^0\rangle=\langle\eta_3^0\rangle=0$. L is explicitly violated and again, non zero masses for neutrinos can be generated by quantum effects.\\
(4) Again, $V_{VL}^\prime\neq 0,\; \langle\chi_2^0\rangle\neq 0$ and/or $\langle\eta_3^0\rangle\neq 0$ is also possible, leading to a phenomenology with the presence of a light pseudo Goldstone Majoron.

\section{\label{sec:sec5}The Neutral Sector}
To present the kind of analysis we are aimed to, let us concentrate on Model D to start with.

The lepton fields for this particular model are included in the structure $S_1+S_2+S_6$ which contains 21 two component spinors, including seven neutral Weyl states. Let us write them in the following way:
\begin{eqnarray*}
\psi_{1L}&=&(\nu_1,l_1^-,E_0^-)_L\sim (1,3,-2/3),\\ 
l_{1L}^+&\sim& (1,1,1),\;\;
E_{0L}^+\sim (1,1,1)\\
\psi_{2L}&=&(l_2^-,\nu_2,N_0^0)_L\sim (1,3^*,-1/3),\;\; l_{2L}^+\sim (1,1,1),\\
\psi_{3L}&=&(\nu_3,l_3^-,E_1^-)_L\sim (1,3,-2/3),\\ 
l_{3L}^+&\sim& (1,1,1),\;\; E_{1L}^+\sim (1,1,1)\\
\psi_{4L}&=&(E_2^+,N_{1}^0,N_{2}^0)_L\sim (1,3,1/3),\\
\psi_{5L}&=&(N_{3}^0,E_2^-,E_3^-)_L\sim (1,3,-2/3),\;\; E_{3L}^+\sim (1,1,1),
\end{eqnarray*}
with the 3-3-1 quantum numbers given in parenthesis.\\

Using the scalars of (\ref{higg}) with the VEV as stated, the mass matrix for the neutral sector in the basis 
$(\nu_1,\nu_2,\nu_3,N_0^0,N_{1}^0,N_{2}^0,N_{3}^0)$ is now of the form:
   
\begin{equation}
M_n=\left(\begin{array}{ccccccc}
0 & 0 & 0 & 0  & A  & -a  & 0 \\
0 & 0 & 0 & 0 & M & 0 & 0\\
0 & 0  & 0  & 0 & B & -b & 0  \\
0 & 0 & 0 & 0 & 0 & M & 0 \\
A & M & B & 0 & 0 & 0 & G \\
-a & 0 & -b & M & 0 & 0 & -d \\
0 & 0 & 0 & 0 & G & -d & 0 
        \end{array} \right),
\end{equation}
where the $M$ value is related to a GUT mass scale coming from the bare mass term $\psi_{2L}C\psi_{4L}+ h.c$; $A,B$ and $C$ are mass terms at the TeV scale $V$, and $a,b$ and $c$ are mass terms at the electroweak scale $v\sim v_1\sim v_2$. The diagonalization of the former mass matrix produces two Dirac massive spinors with masses at the GUT scale and three Weyl massless states that we can associate with the detected solar and atmospheric oscillating neutrinos.

So, up to this point the model has the potential to be consistent with the neutrino phenomenology. But the question is if the three Weyl states remain massless or if they may pick up small radiative masses in the context of the model, or a simple extension of it, something out of the reach of the analysis presented here.

\subsection{\label{sec:sec51}General analysis for 3 families}
Analysis similar to the previous one have been carried through for the neutral fermion sector of the eight anomaly-free lepton structures enumerated in Sec.~\ref{sec:sec4}. The results are presented in Table II.

\begin{table}[here!]{{TABLE II : Tree level neutrinos sectors  }}\label{table2}

\begin{tabular}{|c|c|c|c|}\hline\hline
Model & Number of Weyl  & Massless & Dirac States \\
          & neutral states  & Weyl states   &  at the EW scale    \\ \hline
A:& 6  & 2 & 2 \\
B:& 3  & 3 & 0\\
C:& 8  & 0 & 3\\
D:& 7  & 3 & 0\\
E:& 14 & 0 & 3\\
F:& 13 & 0 & 1\\
G:& 12 & 0 & 3\\
H:& 15 & 0 & 4\\
\hline\hline
\end{tabular} 
\end{table}
According to this Table, only models B and D fulfill the natural condition of having 3 tree-level zero mass neutrinos, which may pick up non zero masses via radiative corrections, with or without the adition of new ingredients. Some other structures may become realistic if new fields are added, and/or if some Yukawa coupling constants are fine tuned to very small values, and/or if discreet symmetries which forbids Yukawa terms are imposed, etc..

Let us see this in the following example.

\subsection{\label{sec:sec52}The 3-3-1 model with exotic electrons}
To see what kind of new ingredients are needed in order to provide masses to the neutral fields in these 3-3-1 models without exotic electric charges, let us briefly view the situation for model B which was introduced in the literature for the first time in Ref.~\cite{ozer}. The neutral fermion sector for this model has been studied in some detail in Refs.~\cite{zapa,sala}, but the approach here is simpler.

The anomaly free fermion structure for this model is~\cite{ozer}:
\begin{eqnarray*}
\psi_{lL}^T&=& (\nu_l^0,l^-,E_l^-)_L\sim (1,3,-2/3),\\ 
l_L^+&\sim& (1,1,1),\;\; E_{lL}^+\sim (1,1,1),\\ 
Q_{iL}^T&=& (d_i,u_i,U_i)_L\sim (3,3^*,1/3),\\
Q_{3L}^T&=&(u_3,d_3,D)\sim (3,3,0),\\
u_{aL}^c&\sim& (3^*,1,-2/3),\;\; U_{iL}^c\sim (3^*,1,-2/3), \\ 
d_{aL}^c&\sim& (3^*,1,1/3),\; \;D_{L}^c\sim(3^*,1,1/3), 
\end{eqnarray*} 
where $l=e,\mu,\tau$ is a lepton family index, $E_{l}^-$ stands for three exotic electron fields, $i=1,2$ for the first two quark families, $a=1,2,3$ is again the quark family index, and there are two exotic quarks with 
electric charge $2/3\; (U_i)$ and other one with electric charge $-1/3\; (D)$. This model does not contain  right handed neutrino fields.

The gauge boson and scalar sectors for this model are exactly the same ones that for the model with right 
handed neutrinos~\cite{long}; but the big differences are that now, the lepton number L is a good quantum number of the model and the gauge bosons does not carry 
lepton number at all, neither the exotic quarks. The scalars $(\eta,\rho,\chi)$ introduced have also L=0, the lepton number cannot be broken spontaneously and, as a consecuence, the neutrinos remain massless even with the inclussion of the radiative corrections.

In what follows and in order to simplify matters and make this model more predictable, 
we consider only the set of two scalar triplets $\chi$ and $\rho$ instead of the set of three 
triplets proposed in the original paper~\cite{ozer}, or the much more complex structure introduced in Ref.~\cite{zapa}. Also, let us take the VEV to be $\langle\chi\rangle^T=(0,v,V)$ and $\langle\rho\rangle^T=(v_1,0,0)$ (as 
in the Economical 3-3-1 model~\cite{ponce} which is going to be studied next). The Yukawa couplings of the leptons to this scalars are now 
\begin{equation}\label{mslo}
{\cal L}^{l}_2= \sum_{l,l^\prime}[(\psi_{lL}^T.\chi)C(h_{ll^\prime}^{e} l_L^{\prime +} 
+h_{ll^\prime}^{E}E^+_{l^\prime L})] + h.c.,
\end{equation}
which for $l,l^\prime=e,\mu,\tau$ saturates all the entries of the $6\times 6$ charged lepton mass matrix and allows tree-level masses only for charged leptons, eventhough there are in (\ref{mslo}) external legs with neutrino fields of the form $\nu^0_{lL}\chi_1^-C(h_{ll'}^el^{'+}_L + h_{ll'}^EE^{+}_{l'L})+ h.c..$  
The possible inclusion of the scalar $\eta$ does not change this situation at all.

Masses for neutrinos can be obtained only by enlarging the model. For this purpose one can introduce a new scalar triplet $\phi=(\phi_1^{++},\; \phi_2^{+},\;\phi_3^+)\sim (1,3,4/3)$ which 
couples to the spin 1/2 leptons via a term in the Lagrangian of the form
\begin{eqnarray}\label{massn}
{\cal L}^l_3&=&\epsilon_{nmp}\sum_{l,l^\prime}h_{ll^\prime}^\nu\phi^n\psi_{lL}^mC\psi_{l^\prime L}^p + h.c.,\\ \nonumber
&=&\sum_{ll^\prime}h^\nu_{ll^\prime}[\phi_1^{++}(l^-_LE^-_{l^\prime L}-l^{\prime -}_LE^-_{lL}) \\ \nonumber 
&+& \phi_2^{+}(E^-_{lL}\nu_{l^\prime L}-E^-_{l^\prime L}\nu_{lL})
+\phi_3^+(\nu_{lL}l^{\prime -}_L-\nu_{l^{\prime}L}l^-_L)] + h.c.,
\end{eqnarray}
which implies lepton number values L$(\phi_1^{++},\; \phi_2^{+},\;\phi_3^+)=-2$ in order to have it conserved  in ${\cal L}^l_3$. Notice that the expression above also provides several external legs with neutrino fields which can be used to generate masses to the neutral fermions via quantum effects. 

\begin{figure}
\centering
\includegraphics{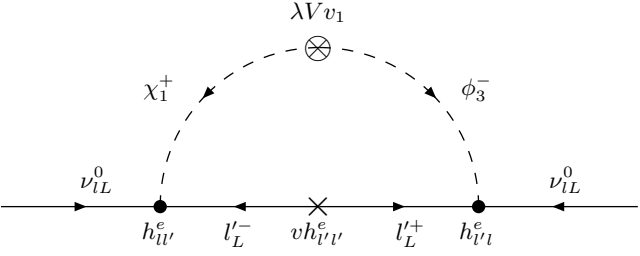}
\caption{\label{fig1}Generation of the neutrino masses via the one loop radiative mechanism in the 3-3-1 model with exotic electrons.}
\end{figure}

Since $\langle\phi\rangle=(0,0,0)$, the new scalar fields are not able to break spontaneously the 
lepton number. But the point is that the lepton symmetry is now explicitly broken in the Lagrangian by a term in the scalar potential of the form 
$\lambda(\phi.\chi)(\rho^*.\chi)$ which violates lepton number by two units and turns on the Zee radiative mechanism in the context of this 3-3-1 model with exotic electrons. As a matter of fact, all the previous ingredients allow us to draw the diagram in Fig.~\ref{fig1} in the context of the field structure presented so far.

Although the scalar sector has three independent fields $(\chi,\rho,\phi)$, its VEV structure is simpler that the one proposed in the original paper~\cite{ozer}.

Neutrino masses in the context of the model analyzed in this section, were studied for the first time in Ref.~\cite{kita}. The main difference between that paper and this one is that in Ref.~\cite{kita}, and in order to implement the Zee-Babu mechanism~\cite{zee, babu} for generating neutrino mass terms, a double charged Higgs scalar $SU(3)_L$ singlet $k^{++}\sim (1,1,2)$ was used instead of our $\phi$ scalar triplet, which is the new and main ingredient of our analysis. So, both papers address to the same problem from two different points of view.

\section{\label{sec:sec6}The economical 3-3-1 model}
The model was introduced for the first time~\cite{ponce,lon} and the quark and lepton content corresponds to the 3-3-1 model with right-handed neutrinos presented above; but the scalar sector is modified, becoming minimal in the sense that only two scalar triplets (with a modified VEV structure) are used in order to break the symmetry. They are:
\begin{eqnarray*}
\rho^T&=&(\rho_1^0,\rho_2^+,\rho_3^+)\sim (1,3^*,2/3), \\
\chi^T&=&(\chi_1^-,\chi_2^0,\chi_ 3^0)\sim  (1,3^*,-1/3), 
\end{eqnarray*}
with the VEV given now by: $\langle\rho\rangle^T=(v_1,0,0)$, and $\langle\chi\rangle^T=(0,v,V)$. 
 
The lepton number L and the global symmetry ${\cal L}$ are as given 
for the model with right-handed neutrinos, and Eq.(\ref{lll}) and the lepton number assignment in (\ref{lnass}) still holds. 

This model has been the subject of several recent studies~\cite{lon,dong2,dong3} and it has the peculiarity that the lepton number L is spontaneously broken due to the fact that L$(\chi_2^0)=2$.

Since the scalar sector is very simple now, the model is highly predictable. As a matter of fact,
 the full scalar potential consist only of the following six terms~\cite{ponce}:
\begin{eqnarray}\nonumber
V(\chi,\rho)&=&\mu_1^2|\chi|^2+\mu_2^2|\rho|^2+\kappa_1|\chi^\dagger\chi|^2
+\kappa_2|\rho^\dagger\rho|^2\\ 
&+&\kappa_3|\chi|^2|\rho|^2+\kappa_4|\chi^\dagger\rho|^2+h.c..
\end{eqnarray}

A simple calculation shows that both, ${\cal L}$ and the lepton number L are conserved by 
$V(\chi,\rho)$ and also by the full Lagrangian, except for some of the following Yukawa interactions which induce masses for the fermions
\begin{eqnarray}\label{yukas}\nonumber
{\cal L}^Y&=&{\cal L}^Y_{LNC}+{\cal L}^Y_{LNV} \\ \nonumber
{\cal L}^Y_{LNC}&=&h^U\chi^*Q_{3L}CU_L^c+h^D_{ij}\chi Q_{iL}CD^c_{jL}\\ 
&&+h^d_a\rho^*Q_{3L}Cd^c_{aL}+h^u_{ia}\rho Q_{iL}Cu_{aL}^c \\ \nonumber
&&+h_{ll^\prime}^e\rho^*\psi_{lL}Cl^{\prime +}_L+h_{ll^\prime}\rho
\psi_{lL}C\psi_{l^\prime L} +h.c.\\ \label{yukasp}
{\cal L}^Y_{LNV}&=& h_a^u\chi^*Q_{3L}Cu_{aL}^c + h^d_{ia}\chi Q_{iL}Cd_{aL}^c \\ \nonumber
&&+h_i^D\rho^*Q_{3L}CD_{iL}^c + h^U_{i}\rho Q_{iL}CU_{L}^c +h.c.,\\ \nonumber
\end{eqnarray}
where the subscripts LNC and LNV indicates lepton number conserving and lepton number violating term respectively. As a matter of fact, 
${\cal L}^Y_{LNV}$ violates explicitly ${\cal L}$ and L by two units.

After spontaneous breaking of the gauge symmetry the scalar potential develops the following lepton number violating terms: 
\begin{eqnarray}\label{vlnv}\nonumber
V_{LNV}&=&
v[\sqrt{2}H_\chi(\kappa_1|\chi|^2+\kappa_3|\rho|^2)] \\
&&+v\kappa_4[\rho_ 1^-(\chi^\dagger\rho)+\rho_1^+(\rho^\dagger\chi)],
\end{eqnarray}
where we have defined as usual $\chi_2^0=v+(H_\chi+iA_\chi)/\sqrt{2}$. $H_\chi$ and $A_\chi$ are the so called CP even and CP odd (scalar and pseudo scalar) components, and for simplicity we are taking real VEV (CP violation through the scalar exchange has not been considered here).

Notice that the lepton number violating part in~(\ref{vlnv}) is trilinear in the scalar fields, and as expected, $V_{LNV}=0$ for $v=0$. From the former expression we can identify $A_\chi$ as the only candidate for a Majoron in this model.

The minimization of the scalar potential has been done in full detail in Ref.~\cite{ponce} (reproduced also in the second paper in Ref.~\cite{lon}). For that purpose two more definitions were introduced: 
$\rho_1^0=v_1+(H_\rho+iA_\rho)/\sqrt{2}$ and 
$\chi_3^0=V+(H^\prime_\chi+iA^\prime_\chi)/\sqrt{2}$. An outline of the main results in Ref.~\cite{ponce}, important for our present discussion,  are:
\begin{itemize}
\item The three CP odd pseudo scalars $A_\chi,\; A^\prime_\chi$ and $A_\rho$, the would be Goldstone bosons, are eaten up by $Z,\; Z^\prime$ and $(K^0+\overline{K}^0)/\sqrt{2}$, the real part of the neutral bi lepton gauge boson.
\item Out of the three CP even scalars, $(vH^\prime_\chi-VH_\chi)/\sqrt{v^2+V^2}$ becomes a would be Goldstone boson eaten up by $i(K^0-\overline{K}^0)/\sqrt{2}$, the imaginary part of the neutral bi lepton gauge boson which picks up L=2 via $H_\chi$. The other two CP even scalars become the SM Higgs boson and one extra Higgs boson with a heavy mass of order $V$ respectively.
\item In the charged scalar sector $(\rho_2^\pm,\; \chi_1^\pm,\; \rho_3^\pm)$ there are four would be Goldstone bosons, two of them are  $(V\chi_2^\pm-v_1\rho_3^\pm)/\sqrt{V^2+v_1^2}$ with L$=\pm 2$, eaten up by $K^\pm$, and other two with L=0 eaten up by $W^\pm$. 
\item Two charged scalars remains as physical states.
\end{itemize}
Counting degrees of freedom tells us that there are in $\chi$ and $\rho$ twelve ones namely: three CP even, three CP odd and six charged ones. Eight of them are eaten up by the eight gauge bosons 
$W^\pm,\; K^\pm, K^0,\; \overline{K}^0,\; Z$, and $Z^\prime$. Four scalars remains as physical states, one of them being the SM Higgs scalar.

Since L is explicitly broken in the context of this model, the most outstanding result in our analysis is that the would be pseudo Goldstone Majoron $A_\chi$, the only CP odd electrically neutral scalar with L=2, has been eaten up by $(K^0+\overline{K}^0)/\sqrt{2}$, the real part of the bi lepton gauge boson. A clever way to avoid an unwanted Majoron!

A variant of this model was considered in Ref.~\cite{gps} where the fermion mass spectrum was studied with the inclusion of a $Z_2$ discrete symmetry which excludes the LNV interactions in the Yukawa potential in equation (\ref{yukasp}). For this variant of the model, ${\cal L}$ is conserved through the entire Lagrangian, the lepton number L is only spontaneous violated by $V_{LNV}$ in Eq.(\ref{vlnv}) and the would be Majoron $A_\chi$ is gauge away, eaten up by $(K^0+\overline{K}^0)/\sqrt{2}$. Notice that being ${\cal L}$ a good quantum number, the spontaneous violation of $SU(3)_L$ implies the spontaneous violation of L via Eq.~(\ref{lll}), something that it is now allowed because the fermion sector for L is vector like and thus non-anomalous.

The economical scalar structure presented here, is not able to reproduce a consistent quark mass spectrum at tree level. By fortune, a careful analysis combining the renormalizable Yukawa interactions in (\ref{yukas}) and (\ref{yukasp}), and the effective dimension-five operators
\begin{eqnarray}\label{yuknr}
{\cal L}_{NR}&=&\frac{\epsilon_{nmp}}{\Lambda}\left[ \chi^n\rho^mQ_{3L}^pC\left({\lambda_3^U}U_L^c+\sum_{a=1}^3{\lambda_a^u}u_{aL}^c\right)\right.\\ \nonumber 
&+&\left.\chi^{*n}\rho^{*m}\sum_{i=1}^2Q_{iL}^pC\left({\lambda_i^d}D_L^c+\sum_{a=1}^3{\lambda_{ia}^d}d_{aL}^c\right)\right],
\end{eqnarray}

are able to remove the zero quark masses. But the implementation of
${\cal L}_{NR}$ in the former expression requires the introduction of new and heavy scalar fields.

But there remains the question of the quantum effects. A careful analysis shows that the conclusion in Ref.~\cite{msv} related with the quark mass matrices is true; that is, the inclusion of all the one-loop diagrams with the proper Yukawa couplings, still leaves the quark mass matrices with determinant equal to zero. So, contrary to what is stated in Refs.~\cite{lon} and
~\cite{don}, the one-loop diagrams are not able by themselves  to provide a consistent mass spectrum for the quarks in the context of this economical model. But it does not mean that there is a remanent $U(1)$ symmetry in the full Lagrangian as it is erroneously stated in Ref.~\cite{msv} (in fact, in Ref.~\cite{don} it is clearly proved that such a $U(1)$ symmetry does not exists at all). The solution to this puzzle and to the controversy raised between Ref. ~\cite{msv} and \cite{don} lies in the two-loop quantum effects which provides with a consistent quark mass spectrum via Babu type mechanisms~\cite{babu}. But this analysis lies outside the scope of this paper and it will be presented elsewhere.

To conclude this section, let us mention that the version of this economical 3-3-1 model developed in the context of the model with right handed neutrinos, can be extended to any one of the eight 3 family models presented in section (\ref{sec:sec4}).

\section{\label{sec:sec7}Conclussions}
The main motivation of our study was to investigate the neutrino mass spectrum in the framework of the local gauge structure $SU(3)_c\otimes SU(3)_L\otimes U(1)_x$.

Summarizing: we have carried out an extensive analysis of the lepton
number symmetry in the context of the best known versions of the 3-3-1
model. It is interesting to remark that in one of these versions,
namely the so called economical model, one explicitly find the
quite unusual situation of the gauging away of the would be Majoron, poviding in this way the
longitudinal polarization component to a now massive gauge field.

This rare but quite unusual mechanism, is related to the fact that the lepton number
generator $L$ is connected with the $\lambda_8$ generator of $SU(3)_L$, 
as shown in Eq. (\ref{lll}).

\section{Conflict of Interests}
The authors declare that there is no conflict of interests regarding the publication of this paper.

{\bf Acknowledgments} We thank Enrico Nardi for a written communication and Vicente Vento for his comments. W.A.P. and R.H.B. thanks the ``Laboratorio de F\'\i sica Te\'orica" from U. de La Plata in  Argentina for a warm hospitality during the initial stages of the work which has been partially supported by ``Sostenibilidad U. de A. 2014'', and ``Centro de Investigaciones del ITM''.

\end{document}